\documentclass[12pt]{iopart}
\pdfoutput=1
\usepackage{iopams}
\usepackage{graphicx}
\usepackage{color}

\begin{document}

\title[On the noise induced by the measurement of the THz electrical current]{On the noise induced by the measurement of the THz electrical current in quantum devices}

\author{D. Marian and X. Oriols}
\address{Departament d\rq{}Enginyeria Electr\`{o}nica, Universitat Aut\`{o}noma de Barcelona, Spain.}
\ead{xavier.oriols@uab.es}
\author{N. Zangh\`i}
\address{Dipartimento di Fisica dell'Universit\`a di Genova and INFN sezione di Genova, Italy}

\begin{abstract}
From a quantum point of view, it is mandatory to include the measurement process when predicting the time-evolution of a quantum system. In this paper, a model to treat the measurement of the (TeraHertz) THz electrical current in quantum devices is presented. The explicit interaction of a quantum system with an external measuring apparatus is analyzed through the unambiguous notion of the Bohmian \emph{conditional wave function}, the wave function of a subsystem. It it shown that such THz quantum measurement process can be modeled as a weak measurement: The systems suffers a small perturbation due to the apparatus, but the current is measured with a great uncertainty. This uncertainty implies that a new source of noise appears at THz frequencies. Numerical (quantum Monte Carlo) experiments are performed confirming the weak character of this measurement. This work also indicates that at low frequencies this noise is negligible and it can be ignored. From  a classical point of view, the origin of this noise  due to the measurement at THz frequencies can be attributed to the plasmonic effect of those electrons at the contacts (by interpreting the contacts themselves as part of the measuring apparatus).  
\end{abstract}


\maketitle

\section{Introduction}
\label{intro}

\emph{What does it mean measuring the electrical current at Terahertz (THz) frequency?} Answering this question is not easy neither from an experimental nor theoretical point of view. At such frequencies, the displacement current (related to time-dependent variations of the electric field) becomes even more important than the conduction current (particles crossing a surface). In general, for semi-classical electron device simulations, it is usually assumed that the interaction with an external measuring apparatus does not alter the properties of the system itself. On the contrary,  for quantum device simulations, it is mandatory to take into account the effect of the apparatus on the measured system. The quantum device evolves differently if the system is measured or not because of the uncertainty principle. 

\begin{figure}[h!!!]
\centering
\includegraphics[width=0.80\columnwidth]{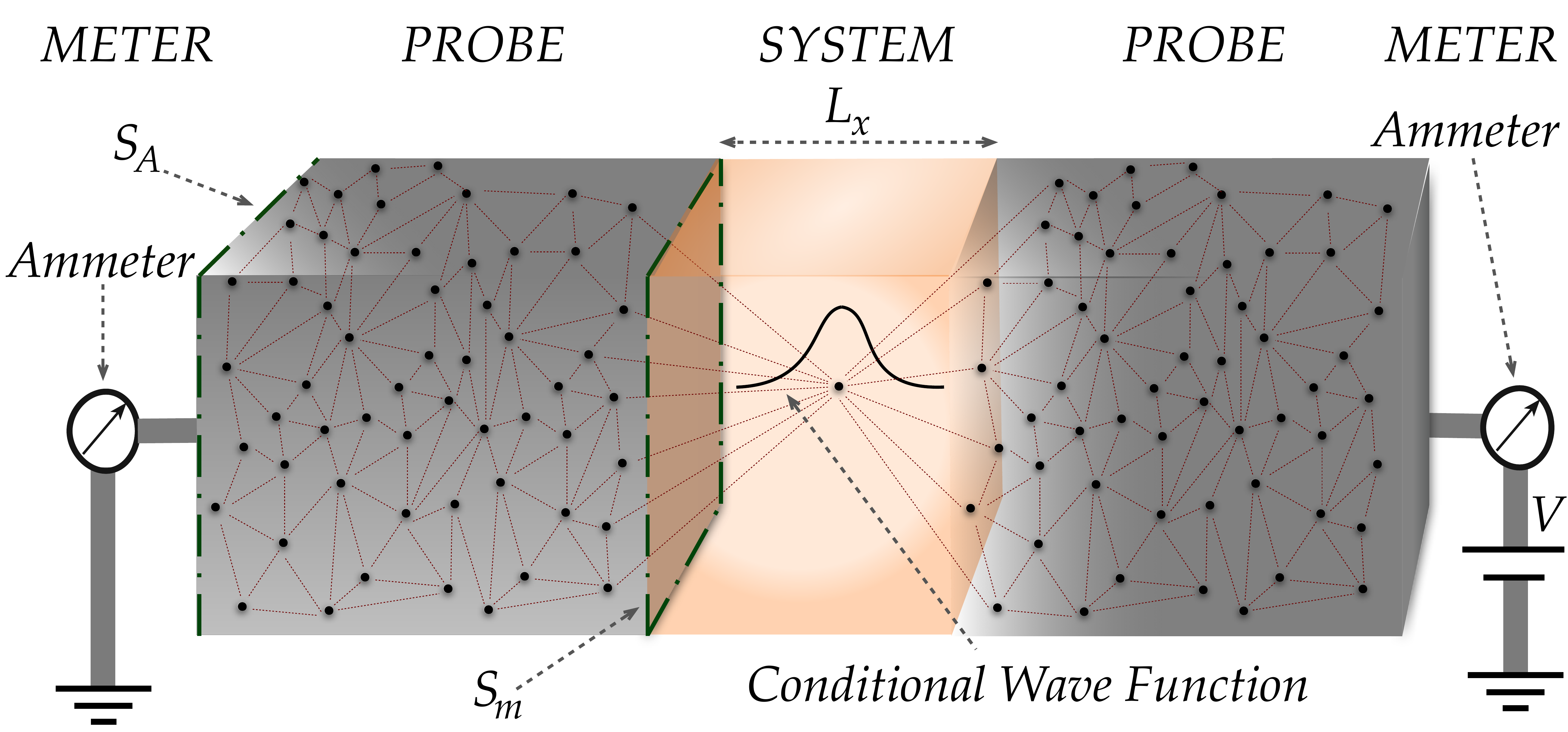}
\caption{Schematic representation of the studied system. We have separated the all problem in three parts, in the middle there is the \emph{system} we are interested in, which interacts through Coulomb interaction (red dashed line) with the electrons in the metal cable (\emph{probe}). Finally the probe interacts with the ammeter (\emph{meter}) which gives the final result of the measurement. The surfaces $S_m$ and $S_A$ (green dashed dotted line) used in the text are indicated.}
\label{figure1}
\end{figure}

In typical quantum device simulations, see \fref{figure1}, the whole setup is divided into the \emph{system} (also known as the active region of the device) of which we want to get informations and the measuring apparatus, composed by \emph{probe} and \emph{meter}, which is responsible to extract the information from the \emph{system}. In principle, one can envision three options for considering the interaction of the system with an external apparatus in quantum device simulations:

\begin{enumerate}
\item The first option is to not consider the measurement apparatus and take directly the informations from the simulated non-measured quantum \emph{system}.
\item The second option is to look for an operator which encapsulates the effect/perturbation of the apparatus on the wave function of the measured \emph{system} and take directly the information from the evolution of the system including the operator in the equation of motion.
\item The third option, which will be investigated in the present paper, is to include the system and apparatus in the simulations and get information from the simulated system+apparatus. We will use the Bohmian trajectories which provides a privilege framework to pursue this option.
\end{enumerate}

Hereafter we elaborate more on the options (i), (ii) and (iii) briefly exposed, trying to underline the advantages and disadvantages of each one. 

\subsection{THz current without modeling the measurement apparatus}
\label{nothing}

The option of not including the apparatus in the simulations seems a very bad choice in order to model the measurement of the electrical current, but actually many electron device simulators are carried out in this way. In fact, as a byproduct of this work, we will show that in the DC regime (direct current), and at very low frequency, it is not necessary to include the apparatus to get accurate values of the current. We give a simple argumentation. We can reasonably assume that an electron device is ergodic, i.e. the mean value obtained from the ensemble is equal to the mean value in time of a certain quantity (the detailed discussion about the ergodic assumption is far from the scope of this paper). This implies that one has to measure the system only once to get the DC current  and it does not matter the subsequent evolution of the measured system. So, in this way the problem of including the measuring apparatus can be avoided, and it is possible to obtain reliable results from the simulate system at DC. However, at THz frequency measurements, there is no proper argument that justifies the non inclusion of the apparatus.   

\subsection{THz current modeling the measurement apparatus with operators}
\label{operator}

This option is based on the traditional quantum mechanical procedure to describe the interaction between the quantum system and the measuring apparatus (the cables, the environment, etc.) by \emph{encapsulating} the latter into a non-unitary operator. This has the great practical advantage of reducing the computational burden of the simulation: only the degrees of freedom of the system are simulated. However, at THz frequencies, many questions about the properties of such operator arise: \emph{which is the operator that determines the (non-unitary) evolution of the wave function when measuring the electrical THz current?} \emph{Is it ``continuous'' or ``instantaneous''? with a ``weak'' or ``strong'' perturbation of the wave function?} \cite{Traversa2013} To the best of our knowledge, no such THz current operator has been presented. So, it seems that this option, although feasible in principle, is not easy to be pursued. 

\subsection{THz current modeling the measurement apparatus with quantum trajectories}
\label{bohm}

In this work, we follow the third option. We will discuss a model for an ammeter that measures the total (conduction plus displacement) current at THz frequency \cite{IWCE,Oriols_2002}. We will consider the interaction between the electrons in a metal surface (\emph{probe}), working as a sensing electrode, and the electrons in the device active region (\emph{system}). In the following we will explain in details the model that we have developed. We will show that the measurement of the electrical current in a large metallic surface implies an unavoidable source of noise \cite{IWCE}, which is generally ignored in most of the high-frequency quantum simulations, and a small perturbation of the quantum system. These two properties, in the context of quantum measurements, mean that the THz measurement of the current can be interpreted as a weak measurement \cite{AAV}.

\section{Model development}
\label{model}

As said in the introduction, it is not easy at all to model the measurement of the THz current \cite{Benali_2012,Albareda_2012}, and before entering into the details of our model let us describe the situation we have to face when addressing this problem. In principle one has to consider the setup depicted in \fref{figure1}, where there is a typical two terminal device contacted by two cables and with two ammeters. Ideally one desires to solve the complete problem quantum mechanically, but unfortunately this is not accessible with the up-to-date computers capabilities. This is the well known many body problem: only $3$, $4$, $5$ degrees of freedom can be treated fully quantum mechanically \cite{Oriols2007}. Thus an approximation, to tackle our measurement problem in THz quantum devices, is required. 

As a first approximation one would desire a method able to tackle as many degrees of freedom as possible and that it is able to provide the output results (in our case the measured current). Our idea is to find an appropriate formalism which is able to include the measuring apparatus (or at least a part of it) in the simulations. For this aim it turns out to be very useful an alternative version of quantum mechanics named Bohmian mechanics \cite{Durr1992,Durr_2013,Bohm_1952,Albareda2013}. This theory uses the standard quantum mechanical wave function, which evolves according to the usual Schr\"odinger equation, and attributes definite positions for the particles at each time. 

Let us briefly review the two basic laws of Bohmian mechanics. The first law says that the many particle wave function is solution of the well-known Schr\"odinger equation: 

\begin{eqnarray}
&&i \hbar \frac{\partial \Psi(\mathbf{x}_1,\mathbf{x}_2,...,\mathbf{x}_N,t)}{\partial t} = \nonumber \\
&=&\left[ -\sum_{i=1}^{N} \frac{\hbar^2}{2m_{i}} \frac{\partial^2}{\partial \mathbf{x}_{i}^2} +V(\mathbf{x}_1,\mathbf{x}_2,...,\mathbf{x}_N,t) \right] \Psi(\mathbf{x}_1,\mathbf{x}_2,...,\mathbf{x}_N,t).
\label{sch}
\end{eqnarray} 

The second law is the \emph{guidance equation} for each particle, which gives its evolution in time:

\begin{eqnarray}
\frac{d \mathbf{X}_{k}(t)}{dt} = \frac{\hbar}{m_{k}} \!Im\! \left( \frac{\nabla_{k} \Psi(\mathbf{x}_1,\mathbf{x}_2,...,\mathbf{x}_N,t)}{\Psi(\mathbf{x}_1,\mathbf{x}_2,...,\mathbf{x}_N,t)} \right) \!\! \Big|_{\mathbf{x}_1=\mathbf{X}_1(t),...,\mathbf{x}_k=\mathbf{X}_k(t),...,\mathbf{x}_N=\mathbf{X}_N(t)},
\label{trajec}
\end{eqnarray}

where we denote the actual positions of the particles with capital letters, i.e. $\mathbf{X}_k(t)$ means a trajectory, while $\mathbf{x}_k$ is a degree of freedom of the problem. 

Along with the many particle wave function in equation \eref{sch} which describes, together with the particle trajectories in equation \eref{trajec}, the theory provides an unambiguous definition of the wave function of a subsystem called \emph{conditional wave function} \cite{Durr1992,Durr_2013,Oriols2007,Norsen2014}. The latter is simply defined from the many particle wave function $\Psi$ where all the degrees of freedom are substituted by the actual position of the particles, i.e. $\mathbf{x}_k \rightarrow \mathbf{X}_k(t)$, except for the particle considered. For example the conditional wave function of particle $1$ is simply given by:

\begin{eqnarray}
\psi_1(\mathbf{x}_1,t) \equiv \Psi(\mathbf{x}_1,\mathbf{X}_2(t),...,\mathbf{X}_N(t),t).
\label{def-cwf}
\end{eqnarray}

The fundamental point that justifies the relevance of the conditional wave function is that the trajectories obtained from the many particle wave function $\Psi$ are exactly the same that the trajectories computed from the conditional wave function $\psi_k$: 

\begin{eqnarray}
\frac{d \mathbf{X}_{k}(t)}{dt} &=& \frac{\hbar}{m_{k}} Im \left( \frac{\nabla_{k} \Psi(\mathbf{x}_1,\mathbf{x}_2,...,\mathbf{x}_N,t)}{\Psi(\mathbf{x}_1,\mathbf{x}_2,...,\mathbf{x}_N,t)} \right) \Big|_{\mathbf{x}_1=\mathbf{X}_1(t),...,\mathbf{x}_k=\mathbf{X}_k(t),...,\mathbf{x}_N=\mathbf{X}_N(t)} \nonumber \\
&\equiv& \frac{\hbar}{m_{k}} Im \left[ \frac{\nabla_{k} \psi_{k}(\mathbf{x}_{k},t)}{\psi_{k}(\mathbf{x}_{k},t)} \right] \Big|_{\mathbf{x}_{k} = \mathbf{X}_k(t)}.
\label{vel-cwf}
\end{eqnarray}

Thus we can obtain the same trajectory, for example for particle $k$, either from the many particle wave function $\Psi$ or from the conditional wave function $\psi_k$. The remarkable fact is that the conditional wave function defined in equation \eref{def-cwf} has its own evolution equation 

\begin{eqnarray}
&&i \hbar \frac{\partial \psi_1(\mathbf{x}_1,t)}{\partial t} = \left[ -\frac{\hbar^2}{2m_1} \frac{\partial^2}{\partial \mathbf{x}_{1}^2} + V(\mathbf{x}_1,\mathbf{X}_2(t),...,\mathbf{X}_N(t),t) \right. \nonumber \\
&+& \left. A(\mathbf{x}_1,\mathbf{X}_2(t),...,\mathbf{X}_N(t),t) +iB(\mathbf{x}_1,\mathbf{X}_2(t),...,\mathbf{X}_N(t),t) \right] \psi_1(\mathbf{x}_1,t),
\label{cwf}
\end{eqnarray}

where $V$ is called \emph{conditional potential}, i.e. the potential felt by particle $1$ because of all the other particles, while the real and imaginary potential $A$ and $B$ are defined from the  many particle wave function $\Psi$ (for a detailed derivation of these potentials see Refs. \cite{Oriols2007,Norsen2014}). Let us mention that the interaction between the quantum system and the measuring apparatus, studied through quantum (Bohmian) trajectories, provides a microscopic definition of the interaction with the apparatus, without the need of postulating an operator \cite{IWCE,Alarcon_2009,Oriols_1998}.  

\subsection{How to calculate the output results}

Before entering in  the details of the model let us mention how the measurement of the electrical current is performed in terms of positions $\mathbf{X}_i(t)$ of the (Bohmian) electrons and conditional wave functions $\psi_i$.  Let us specify that, although we only consider the potential $V$ in equation \eref{sch} (a quasi static approximation), we implicitly assume that the dynamics of electrons are compatible with Maxwell equations so that there is an electromagnetic propagation of the total current along the cable (that connects the quantum system and the ammeter in \fref{figure1}). The total current on $S_m$ is equal to the current on the surface, $S_A$, far from the active region. This equivalence (due to the divergenceless of the total current) is exact for the sum of the particle plus the displacement currents, but not for the particle current alone \cite{Oriols_2005,Oriols_2001}. Once we have considered such propagation, the ammeter transforms the total current $S_A$ into a pointer value \cite{New_2015}. So, as reported in \fref{figure1}, there are three main parts involved in this measurement. First, the \emph{system} or the device active region. Second, the \emph{probe} which is responsible to translate the current until the  \emph{meter}.  And, third, the \emph{meter} itself that actually translate the value of the current into a pointer position in the ammeter. 

The total current, $I_{T}(t) = I_{p}(t)+I_{d}(t)$, is composed by the displacement component $I_d(t)$, defined as the surface integral of the temporal derivative of the electric field, plus the particle component, $I_{p}(t)$, defined as the net number of electron crossing the surface $S_A$ \cite{Alarcon_2009}. For simplicity, we shall focus only on the displacement component of the total current (no electrons crossing the surface when the current is measured). So, $I_{d}(t)$ can be computed as the time derivative of the flux $\Phi$ of the electric field $\mathbf{E} \equiv \mathbf{E}(\mathbf{X}_1(t),...,\mathbf{X}_N(t),t)$ produced by all (system \emph{plus} apparatus) N electrons, described by positions $\mathbf{X}_i(t)$ and conditional wave functions $\psi_i$, on the surface $S_A$ using the relation \cite{IWCE,Albareda2013}:

\begin{equation}
I_{d}(t)  =  \int_{S_A}  \epsilon(\mathbf{r})\frac{d\mathbf{E}}{dt} \cdot d\mathbf{s} =  \sum_{i=1}^{N} \mathbf{\nabla} \Phi(\mathbf{X}_i(t)) \cdot \mathbf{v}_i,
\label{eq-1}
\end{equation}

where the flux $\Phi$ explicitly defined in the appendix depends on each electron position and $\mathbf{v}_i$ is the Bohmian velocity, which is obtained from $\psi_i$ using equation \eref{vel-cwf}.

For a detailed derivation of equation \eref{eq-1} the reader is guided to Ref. \cite{New_2015}. We underline that equation \eref{eq-1} provides the measured current by the ammeter in function only of the position of the particles, $\textbf{X}_i(t)$, and the velocity of the particles $\mathbf{v}_i$. We underline that equation \eref{eq-1} provides the current carried by all the particles (\emph{system}, \emph{probe} and \emph{meter}) and not only the current generated by the particles of the \emph{system}.

In the rest of the paper we focus on the situation in which the electrical current is measured in a large metal surface, large in the sense that the squared distance from the particle to the surface, say $L_x$ of \fref{figure1}, is much greater than the surface where the current is recollected, say $S_A$. We consider also the simplest case where there is only one particle in the active region of the device, say particle 1. In this particular case equation \eref{eq-1} becomes:

\begin{equation}
I_{d}(t) \!\propto\! \frac{d \Phi(\mathbf{E})}{dt} \!=\!  \frac{d}{dt} \left( \alpha  X_1(t) + \sum_{j=2}^{N} \Phi(\mathbf{X}_j(t))  \right)\! \propto  \!v_{x_1}  +   \sum_{j=2}^{N}  \nabla \Phi(\mathbf{X}_j(t)) \cdot \mathbf{v}_j, 
\label{total_current}
\end{equation}

where $v_{x_1}$ is the $x$-component of the Bohmian velocity of particle $1$. These results, equations \eref{eq-1} and \eref{total_current}, have been obtained in Ref. \cite{New_2015}, nevertheless we reported a brief summary in Appendix A. In particular, equation \eref{total_current} will be used extensively in the rest of the text.  We emphasize that equation \eref{total_current} provides a relation between (i)  the total current of the quantum system itself which is proportional to $v_{x_1}$ and (ii) the current effectively measured by the ammeter which we denote by $I_{d}(t)$. Notice that Bohmian mechanics is a quantum theory without observers. Therefore, it permits of talking about the current of the quantum system even though it is not the measured value. 

\subsection{System-Probe interaction}

\begin{figure}[h!!!]
\centering
\includegraphics[width=0.30\columnwidth]{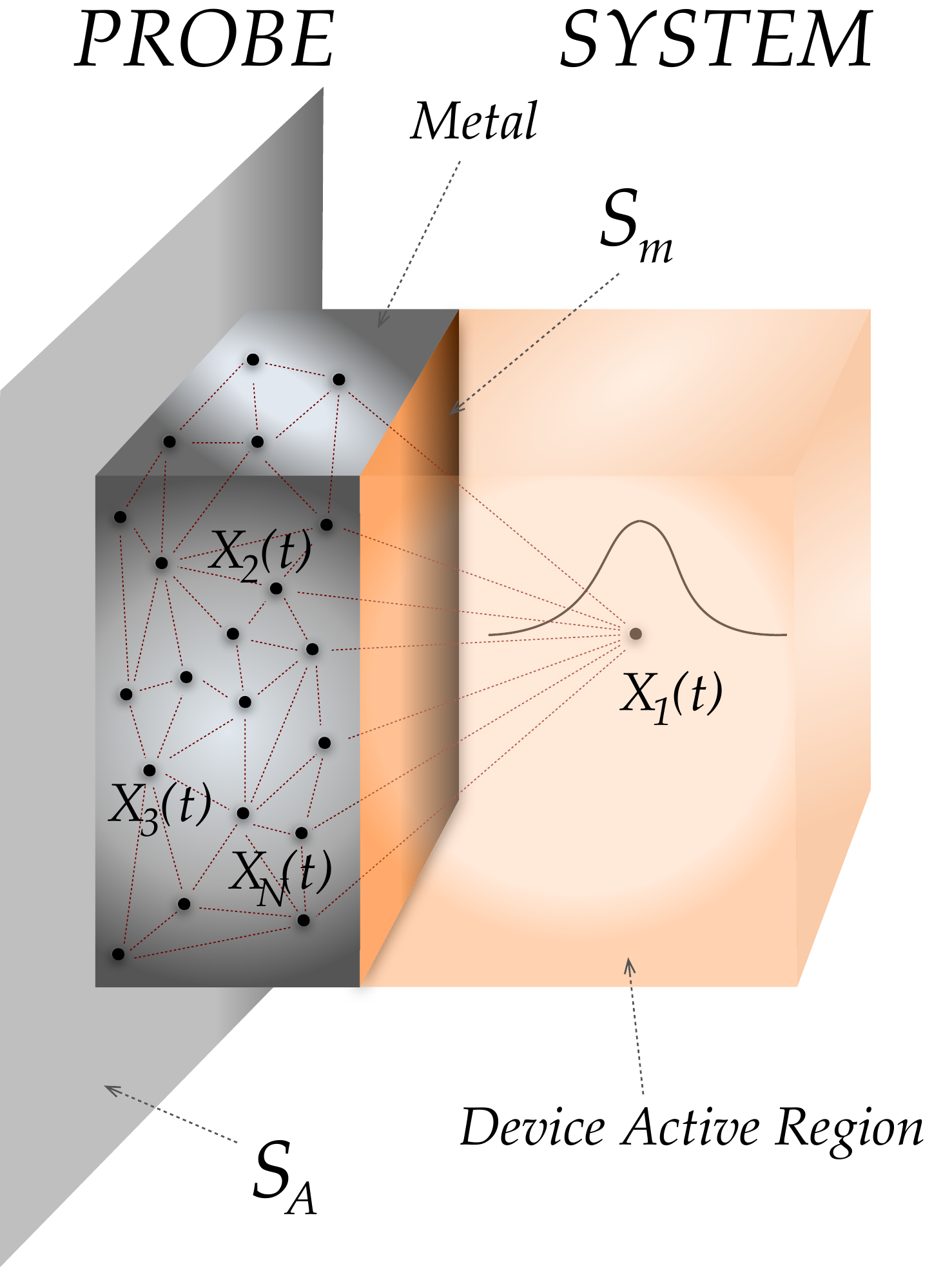}
\caption{It is schematically depicted the coulomb interaction (red dashed lines) and the conditional wave function (black solid line) solution of equation \eref{eq-conditional}. This is the actual system used in the numerical simulations in \sref{num}.}
\label{figure2}
\end{figure}

In principle, we would need to consider all the particles, described by $\mathbf{X}_i(t)$ and $\psi_i$, of \fref{figure1} in order to simulate exactly the whole system, but as said above this is not possible for the well known quantum many body problem \cite{Oriols2007,Traversa_2011}. If we focus on the dynamics of the particle in the active region of the device, with position $\mathbf{X}_1(t)$ and conditional wave function $\psi_1(\mathbf{x}_1,t)$, we easily realize that its interaction with all other particles can be divided according to its effect on $\mathbf{X}_1(t)$. The first type of particles are those $N-1$ particles close to $\mathbf{X}_1(t)$ where the full Coulomb interaction with this particle is relevant. We simulated explicitly how the $N-1$ particles affect $\mathbf{X}_1(t)$ and, very importantly, also how $\mathbf{X}_1(t)$ affect the $N-1$ particles. The second type of particles are those far enough from $\mathbf{X}_1(t)$ so that we assume that they slightly affect $\mathbf{X}_1(t)$ and the effect of this particle on them is negligible. Then, the global effect of these particles on $\mathbf{X}_1(t)$ is computed by the (mean field) quasi-electrostatic boundary conditions. Clearly this distinction between the two types of particles depends on the distance to (and the energies of) the electron in the active region of the device. In summary, we only consider the particle,  described by $\mathbf{X}_1(t)$ and $\psi_1(\mathbf{x}_1,t)$, belonging to the quantum system, and the nearest electrons, described by $\{\mathbf{X}_2(t),\psi_2(\mathbf{x}_2,t)\},...,\{\mathbf{X}_N(t),\psi_N(\mathbf{x}_N,t)\}$, in the metal surface $S_m$ (see \fref{figure2}). The rest of electrons are included in the (quasi-)electrostatic boundary conditions of the problem. In addition, for simplicity, we are considering that the particle in the active region of the device is moving only in the transport direction $\mathbf{x}_1 \equiv (x_1,0,0)$. Thus, we are explicitly neglecting the part of the ammeters (\emph{meter}), where the current is actually measured and we concentrate only on the system-probe interaction.

In principle the exact solution of the conditional wave function is provided by equation \eref{cwf}, but as already briefly explained, the potentials $A$ and $B$ are exactly known only if the many particle wave function $\Psi$ is known. Fortunately, we can provide a suitable approximation for the problem studied here. We can use the approximation, reported in Refs. \cite{Oriols2007,Norsen2014}, where the spatial dependence of the potentials $A$ and $B$ is neglected: $A(x_1,\mathbf{X}_2(t),...,\mathbf{X}_N(t),t) \approx A(X_1(t),\mathbf{X}_2(t),...,\mathbf{X}_N(t),t)$ and $B(x_1,\mathbf{X}_2(t),...,\mathbf{X}_N(t),t) \approx B(X_1(t),\mathbf{X}_2(t),...,\mathbf{X}_N(t),t)$, while keeping the spatial dependence of $V(x_1,\mathbf{X}_2(t),...,\mathbf{X}_N(t),t)$.
Thus the evolution of the conditional wave function of the electron in the device active region is given by the following equation:

\begin{equation}
\label{eq-conditional}
i\hbar \frac{\partial \psi(x_1,t)}{\partial t} = \left[ H_0 + V \right] \psi(x_1,t), 
\end{equation}

where $V = V(x_1,\mathbf{X}_2(t),...,\mathbf{X}_N(t))$ is the conditional Coulomb potential felt by the system and $H_0$ is its free Hamiltonian. Obviously, the electron in the active region of the device is still affected by the other particles composing the metal surface through $V$, this point is crucial for including the back action of the measuring apparatus on the quantum system, i.e. the actual effect of the measuring apparatus on the measured system.

On the other hand, the electrons in the metal surface (the \emph{probe}) are simulated as follows. Each electron $\mathbf{X}_k(t)$ interact with the others electrons in the metal $\mathbf{X}_2(t),...,\mathbf{X}_{k-1}(t),\mathbf{X}_{k+1}(t),...,\mathbf{X}_N(t)$ \emph{plus} with the electron in the active region of the device $\mathbf{X}_1(t)$. The simulations reported hereafter in section \ref{num} are performed considering approximately $1000$ electrons in the metal surfaces. For simulating such trajectories $\mathbf{X}_{k}(t)$ we can take the time derivative of equation \eref{vel-cwf}, obtaining:

\begin{eqnarray}
\frac{d^2\mathbf{X}_k(t)}{dt^2} = -\frac{1}{m_k} \mathbf{\nabla}_k \left( V+Q \right) |_{\mathbf{x}_1=\mathbf{X}_1(t),...,\mathbf{x}_k=\mathbf{X}_k(t),...,\mathbf{x}_N=\mathbf{X}_N(t)},
\label{second}
\end{eqnarray} 

where $Q = -\sum_{i=1}^{N} \frac{\hbar^2}{2m_i} \frac{\nabla_i^2 |\Psi|}{|\Psi|}$ is called quantum potential and where $m_i$ is the electron's mass in the metal. Because of the large number of the electrons, the contribution $\nabla_i^2 |\Psi|$ to the quantum potential is somehow randomize and it becomes small compared to $V$ so we can approximate equation \eref{second} as follows:

\begin{eqnarray}
\frac{d^2\mathbf{X}_k(t)}{dt^2} = \frac{\mathbf{F}_k}{m_k}.
\end{eqnarray}

The force $\mathbf{F}_k$ consists of two contributions, the Coulomb force $\mathbf{F}^{Coulomb}_k=-\mathbf{\nabla}_k V$ and a viscosity term (in order to simulate the interaction with phonons):

\begin{eqnarray}
\mathbf{F}_k = \mathbf{F}^{Coulomb}_k -\gamma \mathbf{v}_k
\label{phonon}
\end{eqnarray}
where $\gamma = 3.374 \cdot 10^{-17} \;\; Kg/s$. A more realistic treatment of the irreversible dynamics due to electron-phonon interaction, beyond expression \eref{phonon}, will certainly provide quantitative (but not qualitative) differences in the results of section \ref{num}. The number of electron in the metal (3D) surface is chosen roughly as the density of Copper ($n_{Cu} = 8.43 \cdot 10^{28} \:\: m^{-3}$). In the simulations reported hereafter, the surface where the electrons in the metal are simulated is $S_m = 2,5 \cdot 10^{-17}\;m^2$ with a width of $5 \cdot 10^{-9}\;m$ and the time step is $\Delta t = 4 \cdot 10^{-17}\;sec$. The type of probe (metal) used in each experiment (number of particles, geometry, etc) obviously affects quantitatively (not qualitatively) the effects reported in \sref{num}. In other words, the measured values depends not only on the system, but also on the type of measuring apparatus.  
In summary, we have presented a microscopic model for studying the interaction between a (measured) quantum system and a measuring apparatus with the Bohmian trajectories formalism (through particle positions $\mathbf{X}_i(t)$ and conditional wave functions $\psi_i(\mathbf{x}_i)$). We have been able to include the back action of the apparatus on the measured system and we have provided an explicit equation to calculate the measured output current. We remark that, in the model just presented, not only the particle $x_1$ in the active region of the device is affected by the electrons in the metal surface, but also vice versa.

\section{Does it exist an unavoidable source of noise due to the measurement of the THz current?}
\label{new-noise}

In the previous section we have developed an equation \eref{total_current} for the measurement of the electrical current at THz frequency plus a model to determine the equation of motion for all the particles $\mathbf{X}_i(t)$, $i=1,...,N$. Now we want to clarify the following question: \emph{Does it exist an unavoidable source of noise due to the measurement of the THz current?} Our answer will be supported by numerical results in next section. Here, we provide some qualitative arguments on the physics of the type of measurement we are explaining in this work. We focus the attention on equation \eref{total_current}, which we report here again:

\begin{equation}
I_{d}(t)  \propto  v_{x_1} + \sum_{j=2}^{N}  \nabla \Phi(\mathbf{X}_j(t)) \cdot \mathbf{v}_j.
\label{total_current_2}
\end{equation}

We observe that the current is composed by two terms, the first term is proportional to the velocity of the electron in the active region of the device, $v_{x_1}$. This corresponds to the signal that we want to get from the measured system. The second term in equation \eref{total_current_2}, which depends on all the rest of electrons composing the probe, provides the noise of the measurement process. We can see that the instantaneous current, $I_{d}(t)$,  in equation \eref{total_current_2} is affected by this term. In fact the random movement of the electrons in the probe produces a random current output. So, we can interpret equation \eref{total_current_2} as the sum of the signal, first term, plus the \emph{additional source of noise}, the second term. The signal to noise ratio will be discussed in the next section. But is it actually a \emph{new} source of noise? This movement of the electrons are also known as \emph{plasmons}, i.e. collective motions of the electrons composing the metal surface. We assume that the contacts (responsible for the transmission of the current from the quantum system to the ammeter) are an unavoidable part of any measuring ammeter. 

The question here is if this additional source of noise has to be taken into account when performing quantum device simulations or not. We can step back to the three options we have enumerated in the introduction. Option \ref{nothing}, i.e. not modeling the measurement apparatus, needs obviously to include this source of noise because option \ref{nothing} alone  does not contains any information of the apparatus. Instead, working with option \ref{operator}, one must take care that the operator chosen for the measurement of the electrical current at THz frequencies fits with this source of noise we have found. On the other hand, option \ref{bohm}, which is the one used in the present paper, includes naturally this additional source of noise. In addition, the novelty here is the way we have achieved this result: the model we have proposed in the previous section, through the notion of the conditional wave function and Bohmian trajectories, has permitted to interpret, in a still pure quantum mechanical way, the interaction between the measured \emph{system} and the \emph{probe}. So, thanks to the theory and the model, we have been able to deduce this additional and unavoidable source of noise. As a byproduct of this work, we will also obtain that in the DC regime (direct current), and at very low frequency, it is not necessary to include the apparatus (or its noise) to get accurate values of the DC current. In the next section we provide some numerical results supporting the argument just exposed here.

\section{Numerical results}
\label{num}

Here we analyze the results obtained from numerical experiments performed with the model presented in \sref{model}. In particular in \sref{num-1} we show how the measurement of the electrical current in a large surface at THz frequencies provides an additional source of noise. In \sref{num-2} we show how this noise can be interpreted as a weak measurement. The reader can found in Ref. \cite{New_2015} how this weak measurement can be used to reconstruct the Bohmian trajectory of an electron in a multi terminal device. In \sref{num-2} it is also briefly addressed the dependency on frequency of the presented model.

\subsection{Additional source of noise}
\label{num-1}

\begin{figure}[h!!!]
\centering
\includegraphics[width=0.60\columnwidth]{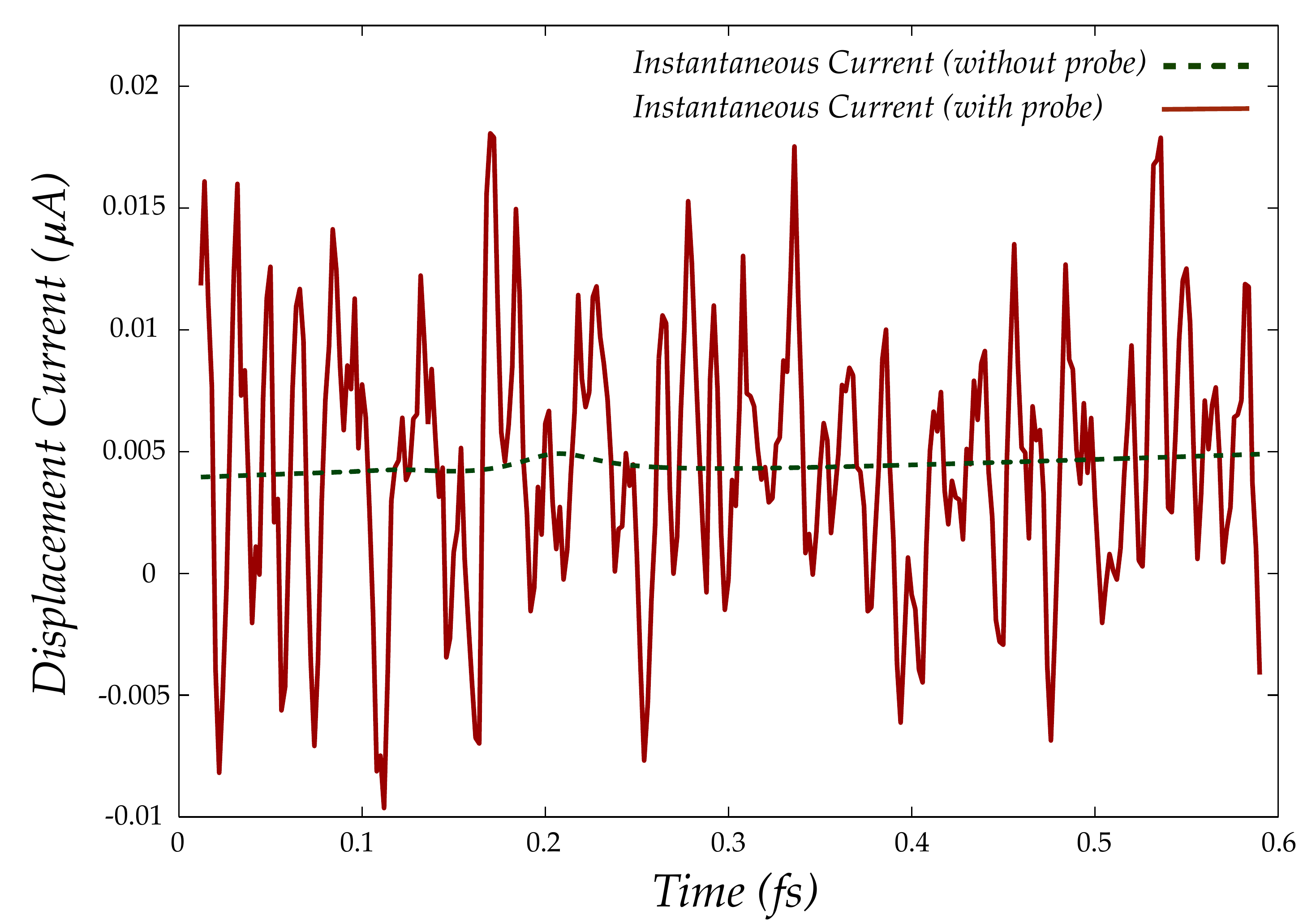}
\caption{Value of the displacement current. With solid line the instantaneous value of the displacement current calculated from equation \eref{eq-1} (with probe) is reported and with dashed line the instantaneous value of the displacement current obtained when considering only particle $x_1$ alone (without probe).}
\label{figure3}
\end{figure}

In \fref{figure3}, we report the instantaneous value of the displacement current measured in the surface $S_A$ when considering all the electrons of the system and the probe (red solid line), and when considering only the electron of the system (green dashed line). 

One can see that the instantaneous current  calculated from equation \eref{total_current_2}, i.e. when considering the contribution of all the electrons in the metal (including the \emph{probe}), differs considerably from the instantaneous current when considering only the electron in the device active region, i.e. without including the \emph{probe}. The difference is due to the second term appearing in equation \eref{total_current_2}. As already discussed in \sref{new-noise}, the random movement of the electrons in the probe produces a random current output, as it can be clearly seen in \fref{figure3}.

The large fluctuations in the instantaneous value of the displacement current reported in \fref{figure3}, when considering the system and the probe, means an additional source of noise due also to the interaction of the electrons in the metal with the particle $x_1$ in the active region of the device.  

\subsection{Weak measurement}
\label{num-2}

\begin{figure}[h!!!]
\centering
\includegraphics[width=0.60\columnwidth]{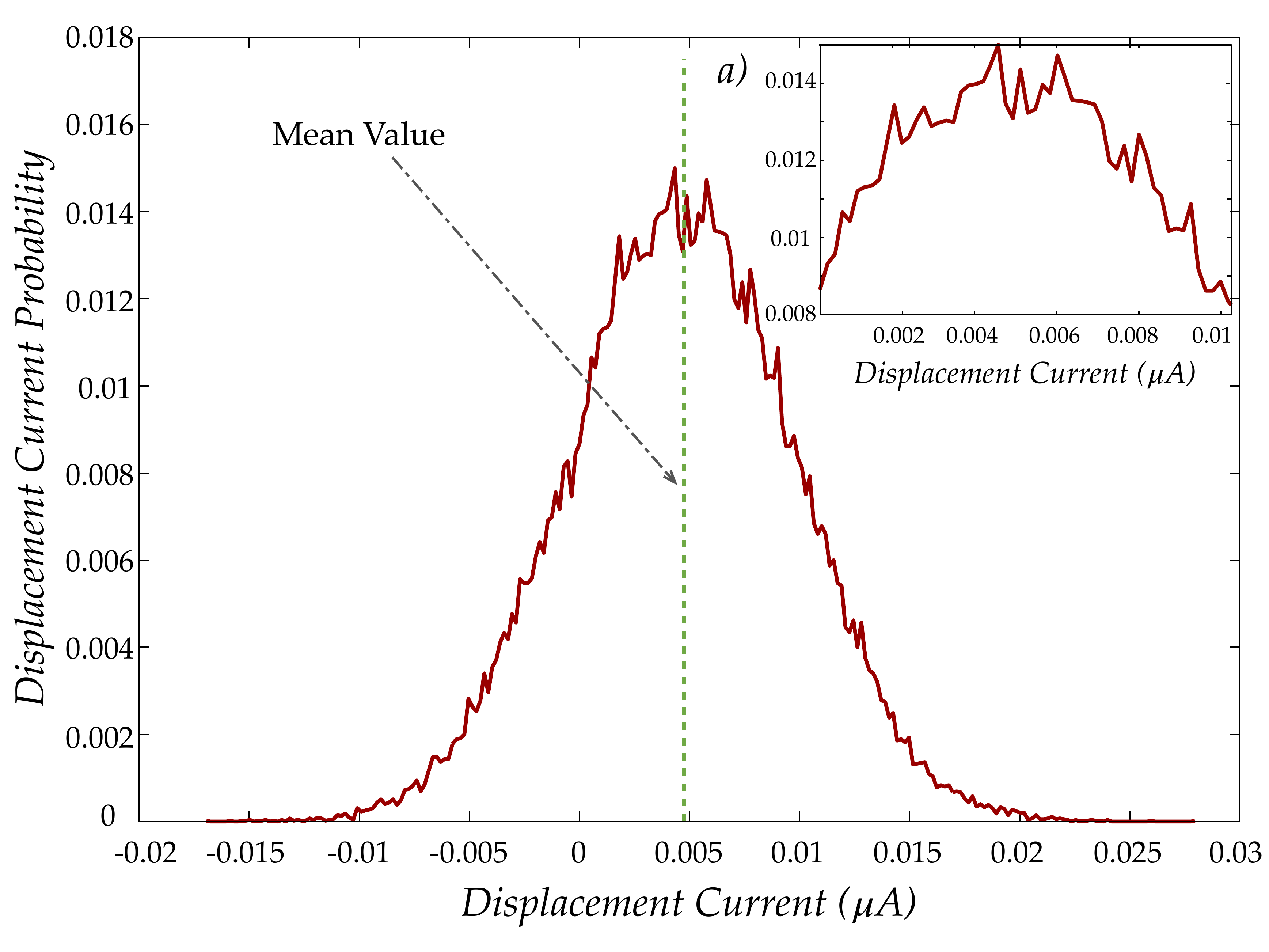}
\caption{Red solid line probability distribution of the measured displacement current from equation \eref{eq-1} from 55000 experiments. Green dashed line mean value obtained without considering the electrons in the \emph{probe}. a) Zoom of the main figure for probability distribution of the displacement current for the values $0\;\mu A$ till $0.01\;\mu A$ around the mean value. }
\label{figure4}
\end{figure}

From the numerical simulations, reported in \fref{figure4}, we observe that the second term in equation \eref{total_current_2} has large fluctuations but it is constant  when evaluated over an ensemble of identically prepared experiments, $\sum_{j=2}^{N} \langle \nabla \Phi(\mathbf{X}_j(t))  \cdot \mathbf{v}_j \rangle\approx const$. So, it is possible to write the ensemble value of equation \eref{total_current_2} as:

\begin{equation}
\langle I_{d}(t) \rangle_{S_A} \propto  \langle p_{x_1} \rangle.
\label{total_main_text}
\end{equation}

Equation \eref{total_main_text} shows that the mean value of the total electrical current in a large metallic surface is proportional to the mean value of the momentum ($x$-component, i.e. the component perpendicular to the surface) of the quantum particle in the device.  In \fref{figure4} it is reported the mean value of the (weak) measured total current computed from equation \eref{total_current_2} which is equal to the value obtained without considering the ammeter, confirming thus equation \eref{total_main_text}.

In addition, see Refs.  \cite{IWCE,New_2015}, the measurement scheme just presented implies also that when the information of the measured current is very noisy, the quantum system is only slightly perturbed, and vice versa. This fact is completely in agreement with the fundamental properties of quantum measurement: if one looks for precise information, one has to pay the price of perturbing the system significantly (the so-called collapse of the wave function or strong measurement). On the other hand if one does not require such a precise information (e.g. the instantaneous value of the displacement current seen in \fref{figure3}) one can leave the wave function of the system almost unaltered (known in the literature as weak measurement \cite{AAV}). 

So, in the context of quantum measurements, the measurement scheme of the total current in a large surface just presented in this work is a \emph{weak measurement} \cite{AAV}. Thus adding the result obtained in equation \eref{total_main_text}, the weak measurement of the total current seems it can be approximated, in the language of Gaussian measurement Kraus operators, by:

\begin{eqnarray}
\hat{I}_{w} = C_w \int dp e^{-\frac{(p-p_w)^2}{2\sigma_w}} |p\rangle \langle p|,
\label{weak-def-I} 
\end{eqnarray}

where $p$ is the momentum (x-component) of the particle in the device and $C_w$ is a suitable constant. We underline that the Gaussian operator defined in equation \eref{weak-def-I} works as an approximation for the output results (of the displacement current) but does not include properly the effect on the quantum system. In fact the gaussian approximation of the measuring operator does not capture all the fluctuations reported in the inset of \fref{figure4} a).  In fact, the apparently random oscillations of the inset  of \fref{figure4} a) really show the difficulties, discussed in option \ref{operator} in the introduction, in properly developing a quantum operator for such type of measurement (small variations on the oscillatory behavior of the operator imply dramatic changes in the dynamics of the system wave packet).

\begin{figure}[h!!!]
\centering
\includegraphics[width=0.60\columnwidth]{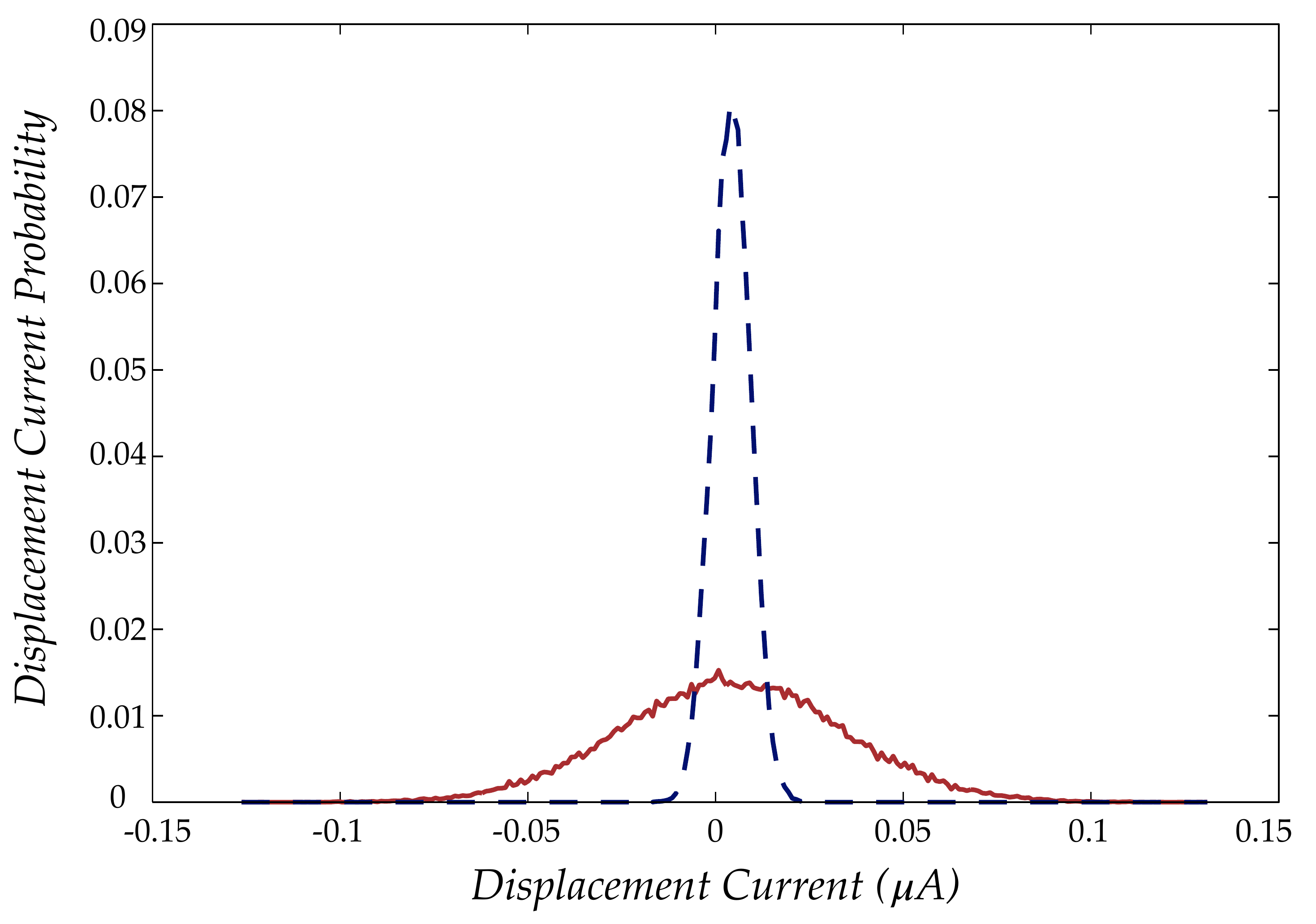}
\caption{Red solid line probability distribution of the measured displacement current at a frequency of $f=500\;\;THz$. Blue dashed line probability distribution of the measured displacement current at $f=50\;\;THz$.}
\label{figure5}
\end{figure}

Finally, we discuss how the width $\sigma_w$ of the weak measurement changes with the frequency. It has been seen that depending on the frequency, the information about the measured total current changes (see Ref. \cite{IWCE}). In \fref{figure5} it is reported how $\sigma_w$ varies with the frequency of the measurement. It can be seen that lowering the frequency yields more precise information about the system (the width of the gaussian decreases). In this sense the source of noise described in this article is unavoidable at high-frequency (THz) regimes, while for lower frequencies, the option \ref{nothing} of not including the apparatus in the simulator still works because such noise become negligible when integrated over a large time interval to provide the DC value.

\section{Conclusions and Discussions}

We have presented a novel model to include how the measuring apparatus affects the value of the measured current at high frequencies (THz) in quantum devices. In particular, 
we have studied the effect of the collective motion of electrons in the metals (the contacts of the device that first collect the total current) on the measurement of the THz electrical current of electronic devices from a quantum point of view. This scheme of the measurement process allows us to differentiate between (i)  the total current of the quantum system itself and (ii) the current effectively measured by the ammeter. According to our analysis, when a large fluctuation in the current appears (i.e. the current of the quantum system is very different from the measured one), the measurement of the THz current implies a slightly perturbation of the quantum system, and vice versa. Additionally, we have also shown that the mean value of the total current, measured by the apparatus obtained by repeating many times the same experiment, provides the strong measured value of the current of the system. Therefore, we conclude: 

\begin{itemize}
\item  The measured current contains an unavoidable source of noise (that corresponds to the plasmonic motion in the contacts) plus a signal (that corresponds to the current of the quantum system itself).
\item The signal-to-noise ratio of the measured current depends (on many physical parameters of the contacts, i.e. on the concrete measuring apparatus, and) on the frequency. For DC values the noise is so small that can be neglected, but for high (THz) frequencies it cannot be dismissed. 
\item In the context of quantum measurements, the THz measurement of the current can be interpreted as a weak measurement.
\end{itemize}

Finally, it should also be mentioned that the weak measurement of the total current at high frequency opens a new path for envisioning experiments for reconstructing (Bohmian) trajectories and wave function of electrons in solid state systems, similar to those already performed for photons \cite{exp1,exp2}. The authors have presented a recent work \cite{New_2015} which pursues this idea.

\section*{Acknowledgement}
The authors acknowledge fruitful discussions with Tom\'as Gonz\'alez, Javier Mateos, Philippe Dollfus and Massimo Macucci. This work has been partially supported by the \lq\lq{}Ministerio de Ciencia e Innovaci\'{o}n\rq\rq{} through the Spanish Project TEC2012-31330,  Generalitat de Catalunya (2014 SGR-384) and by the Grant agreement no: 604391 of the Flagship initiative  \lq\lq{}Graphene-Based Revolutions in ICT and Beyond\rq\rq{}. N.Z. is supported in part by INFN.

\newpage 

\appendix
\section{Displacement current on a large surface}
\label{app}

In this appendix we provide a demonstration of equation \eref{total_current} in the text, the reader can find the same development in Ref. \cite{New_2015}.

According to our discussion in the main text, the capital letters $\{X_i(t),Y_i(t),Z_i(t)\}$ denote the actual (Bohmian) positions of the particles, where $i$ identifies the $i-th$ particle. The flux of the electric field through a general ideal surface $S_A$ in \fref{figure2} (defined as a plane of area  $L_y \cdot L_z$ perpendicular to the $\hat{x}$ direction and placed in $x = x_{A}$, i.e. defined by the points  $\{x_A,0 \leq y' \leq L_y,0 \leq z' \leq L_z\}$, generated by a particle in position $\{X,Y,Z\}$) can be calculated as:

\begin{equation}
\Phi(X,Y,Z)=\int_{S_{A}} \mathbf{E}(X,Y,Z,x_{A},y\rq{},z\rq{}) \cdot d\mathbf{s},
\label{fflux}
\end{equation}

where we have eliminated the subindex $i$ and time $t$ to simplify the notation. The electric field $\mathbf{E}$ is just computed from the Coulomb force of the electron in the mentioned surface. In the simple case in which the particle is located in $\{X,L_y/2,L_z/2\}$, it moves only in the $\hat{x}$ direction and $L_y = L_z \equiv L$  ($S_A = L^2$), equation  \eref{fflux} becomes: 

\begin{eqnarray}
\label{newdisp}
\Phi(X) = \frac{q}{\pi\epsilon}  \tan^{-1}  \left( \frac{S_A}{4(x_A-X)\sqrt{(x_A-X)^2+\frac{S_A}{2}}}\right). 
\end{eqnarray}

Let us evaluate equation \eref{newdisp} in the situation in which $S_A \gg (x_A-X)^2$. This means that the maximum distance (squared) between the electron inside the device active region and  the surface is much smaller than the surface itself. In order to work out an approximate form for equation \eref{newdisp} in this regime, it can be considered the following change of variable $\chi = (x_A-X)$. For simplicity, we assume that the electron is located on the left of the surface (i.e. $X < x_A \rightarrow \chi > 0$), then:  

\begin{eqnarray}
\Phi(\chi)  =  \frac{q}{\pi\epsilon} \tan^{-1}\left( \frac{S_A}{4\chi^2\sqrt{1+\frac{S_A}{2\chi^2}}}\right). 
\label{disp_x}
\end{eqnarray}

Then, calling $\xi^2 = \frac{2\chi^2}{S_A}$, equation \eref{disp_x} becomes

\begin{equation}
\Phi(\xi) = \frac{q}{\pi\epsilon} \tan^{-1} \left( \frac{1}{2\sqrt{\xi^2(1+\xi^2)}}\right),
\label{disp_inter}
\end{equation}

such that the condition $S_A \gg \chi^2$ becomes equivalent to $\xi \ll 1$. So equation \eref{disp_inter} becomes simply:

\begin{equation}
\Phi(\xi)_{\xi^2 \ll 1} =  \frac{q}{\pi\epsilon} \tan^{-1} \left( \frac{1}{2\sqrt{\xi^2}}\right). 
\end{equation}

Remembering that $\tan^{-1}(\alpha \xi)+\tan^{-1}(\frac{1}{\alpha \xi}) = \frac{\pi}{2}$ for $\xi>0$ then one has:

\begin{equation}
\Phi(\xi) =  \frac{q}{\pi\epsilon} \left[\frac{\pi}{2} - \tan^{-1}\left( 2 \xi \right)\right].
\label{disp_chi}
\end{equation}

In equation \eref{disp_chi} the term $\tan^{-1}(2\xi)$ can be expanded obtaining: 

\begin{equation}
\Phi(\xi) = \frac{q}{\pi\epsilon} \left[ \frac{\pi}{2} - 2 \xi + \frac{(2\xi)^3}{3} - ...\right].
\label{disp_taylor}
\end{equation}

This last expression, equation \eref{disp_taylor}, can be truncated at first order of $\xi$ for our large surface. Thus recalling the original variables one arrives at:

\begin{equation}
\Phi(x) = \frac{q}{\pi\epsilon} \left[\frac{\pi}{2} - 2\sqrt{\frac{2}{S_A}} (x_A-X) \right] \propto X.
\label{flux_S}
\end{equation}

Equation \eref{flux_S} is an important result, it demonstrates that the flux of the electric field generated by a particle in a very large surface is proportional to the position of the particle.\\
Now it can be discussed the general problem considered here, i.e. to derive a microscopic analysis of the measurement of the total electrical current in a large metallic surface. In order to do that, one has to ``enlarge'' the system considering also all the electrons composing the metallic surface, as done described in the main text.\\
Without assuming nothing about the dynamics of the electrons in the metal, one can say that they contribute to the flux of the total electric field as described by equation \eref{fflux} by superposition principle. One obtains, suppressing the dependence on $x_A$ and making reference to the position of the electron in the device as $X_1$, the following expression:

\begin{equation}
\label{signal_noise}
\Phi(X_1,\mathbf{X}_2,...,\mathbf{X}_N) = \alpha X_1 + \sum_{j=2}^{N} \Phi(\mathbf{X}_j),
\end{equation}

where the actual Bohmian positions of the particles $\mathbf{X}_k$ have been used and $\alpha$ is a suitable constant.
In equation \eref{signal_noise} one can clearly see that the total electric flux is due to a contribution from the electron in the system $\propto X_1$ and another due to all the other electrons in the metal.\\
So far, it has been considered that the electron in the active region of the device is not crossing the surface and therefore one gets that the total electric current is due only to the displacement current contribution. So, the displacement current becomes: 

\begin{eqnarray}
I_{d} \propto \frac{d \Phi}{dt} =  \frac{d}{dt} \left( \alpha X_1 +   \sum_{j=2}^{N} \Phi(\mathbf{X}_j)  \right) \propto  v_{x_1}  +   \sum_{j=2}^{N}  \nabla \Phi(\mathbf{X}_j) \cdot\mathbf{v}_j, 
\end{eqnarray}

which is the result, equation \eref{total_current}, used in the main text of the article.

\end{document}